\documentclass{amsart}
\usepackage{amssymb, amsmath}

\newcommand{\cH}{\mathcal{H}}
\newcommand{\cL}{\mathcal{L}}

\newcommand{\cE}{\mathcal{E}}
\newcommand{\cW}{\mathcal{W}}
\newcommand{\cP}{\mathcal{P}}

\newcommand{\hH}{\hat{H}}
\newcommand{\hU}{\hat{U}}
\newcommand{\hL}{\hat{L}}

\newcommand{\hcL}{\hat{\cL}}

\newcommand{\hb}{\hat{b}}
\newcommand{\hq}{\hat{q}}
\newcommand{\hr}{\hat{r}} 
\newcommand{\hw}{\hat{w}}
 
\newcommand{\hT}{\hat{T}}
\newcommand{\hW}{\hat{W}}
\newcommand{\hphi}{\hat{\phi}}

\newcommand{\E}{\mathrm{e}}

\newcommand{\Res}{\operatorname{Res}}

\newcommand{\tlambda}{\tilde{\lambda}}
\newcommand{\tf}{\tilde{f}}

\newcommand{\tq}{\tilde{q}}

\newcommand{\rL}{\mathrm{L}}

\newtheorem{theorem}{Theorem} \newtheorem{lemma}{Lemma}

\newtheorem{defi}{Definition}

\begin{document}
\title{Two-step Darboux transformations and exceptional Laguerre polynomials}
\author{David G\'omez-Ullate}
\address{ Departamento de F\'isica Te\'orica II, Universidad Complutense de Madrid, 28040 Madrid, Spain}
\author{ Niky Kamran }
\address{Department of Mathematics and Statistics, McGill University
Montreal, QC, H3A 2K6, Canada}
\author{Robert Milson}
\address{Department of Mathematics and Statistics, Dalhousie University, Halifax, NS, B3H 3J5, Canada}
 \maketitle

\begin{abstract}
It has been recently discovered that exceptional families of Sturm-Liouville orthogonal polynomials exist, that generalize in some sense the classical polynomials of Hermite, Laguerre and Jacobi. In this paper we show how new families of exceptional orthogonal polynomials can be constructed by means of  multiple-step algebraic Darboux transformations. The construction is  illustrated with an example of a 2-step Darboux transformation of the classical Laguerre polynomials, which gives rise to a new  orthogonal polynomial system indexed by two integer parameters. For particular values of these parameters, the classical Laguerre and the type II $X_\ell$-Laguerre polynomials are recovered. 
\end{abstract}  
  
\section{Introduction}
It has recently been shown that the hypotheses of Bochner's
Theorem~\cite{Bo} on the characterization of orthogonal polynomial
systems defined by Sturm-Liouville problems \cite{Lesky} can be
relaxed to give rise to new complete orthogonal polynomial systems. In
some recent papers, the concept of an \textit{exceptional polynomial
  subspace} \cite{GKM3,GKM6} and the closely related notion of
\textit{exceptional orthogonal polynomials} \cite{GKM09a,GKM09b} have
been introduced.  Like the classical examples, exceptional orthogonal
polynomials are eigenfunctions of a second-order differential
operator, but the sequence of eigenfunctions need not contain
polynomials of all degrees, even though the full set of eigenfunctions
still forms a basis of the weighted $\rL^2$ space.

  The first examples of such polynomials were the $X_1$-Jacobi and
  $X_1$-Laguerre polynomials introduced in \cite{GKM09a,GKM09b}.  In
  this co-dimension 1 case, a full characterization of all
  Sturm-Liouville polynomial systems is available thanks to the
  classification of codimension one exceptional polynomial subspaces
  performed in \cite{GKM09a}. The first relation between exceptional
  polynomials and the Darboux transformation was made by Quesne in the
  context of shape invariant potentials in supersymmetric quantum
  mechanics \cite{Quesne1,Quesne2}, where examples of $X_2$
  polynomials are also found. Higher codimension
  families were described for the first time by Sasaki and Odake
  \cite{Sasaki-Odake1} for arbitrary codimension $\ell$. These authors
  also realized \cite{Sasaki-Odake4} that two distinct families of
  $X_\ell$-Laguerre and $X_\ell$-Jacobi polynomials exist, now
  labelled as type I and type II. The existence of precisely two
  families for each class was explained in \cite{GKM10} for the
  Laguerre case, by showing that these families are obtained from the
  classical ones by means of an isospectral algebraic Darboux
  transformation \cite{GKM04a}. Indeed, there are four families of
  algebraic factorizations, only two of which give rise to isospectral
  transformations \cite{GKMDarboux2}. The equivalent result for the
  Jacobi case was shown in \cite{GKM11}, and similar results on the
  relation between the Darboux transformation and exceptional
  polynomials have been published in \cite{STZ10}.

Exceptional orthogonal polynomials and their associated exactly solvable potentials have found numerous recent applications in a range of problems in mathematical physics, such as mass-dependent potentials \cite{Midya-Roy}, Fokker-Planck and Dirac equations \cite{Ho11}, supersymmetric quantum mechanics \cite{DR10, Grandati} or quasi-exact solvability \cite{Tanaka}.

A natural question that arises from the previous works is whether the
isospectral Darboux transformation can be iterated in a Darboux-Crum
form \cite{darboux,crum} in order to obtain families of exceptional
orthogonal polynomials that enjoy spectral characterizations and
completeness properties similar to the ones described above.  Our
purpose in this paper is to give a positive answer to this question by
showing that {\em two-step} isospectral Darboux-Crum transformations
can be used to construct a new set of exceptional orthogonal
polynomials of Laguerre type. This family is indexed by two integer
parameters $(m_1,m_2)$ and a real parameter $k$. The sequence starts
with degree $\ell=m_1+m_2-1$ and it is complete in a weighted
$\rL^{2}$ space endowed with a non-singular weight.  Particular
choices of $m_1$ and $m_2$ correspond to the type II $X_\ell$-Laguerre
and to the classical Laguerre polynomials.
 
Our paper is organized as follows. In Section \ref{Sect2}, we define
the two-step exceptional Laguerre polynomials, present their
characterization as the unique polynomial solutions of a second-order
linear differential equation (Theorem \ref{prop:hLeqn}) and state
their completeness properties (Theorem \ref{prop:2steportho}). We
devote Section \ref{RatFact} to the proof of Theorem \ref{prop:hLeqn},
using rational factorizations, an isospectral Darboux-Crum
transformation and the higher-order intertwining relation
(\ref{higherint}). Section \ref{completeness} addresses then the
completeness property stated in Theorem \ref{prop:2steportho} through
a polynomial approximation argument for weighted $\rL^2$ spaces.
  
 \section{Two-step exceptional Laguerre polynomials}\label{Sect2}
Our goal in this section is to introduce the two-step exceptional Laguerre polynomials, present their relation to second-order linear differential operators and state their completeness and orthogonality properties. The proofs of these results will be given in Sections \ref{RatFact} and \ref{completeness}.

We begin by recalling that for any given real number $k$, the classical
$n$th degree Laguerre polynomial $L_n(z) = L^{(k)}_{n}(z)$ is the
unique polynomial solution of the second-order differential equation
\begin{equation}
  \label{eq:Lneigenrelation}
  \cL_k[L_{n}]=-nL_{n}, 
\end{equation}
where
\begin{equation}\label{Lagk}
 \cL_k[y]=zy''+(k+1-z)y',
\end{equation}
normalized by the condition
\[ \quad L_{n}(z)=\frac{(-z)^n}{n!}+\text{ lower degree terms.}\]  
For $k>-1$, the sequence $\{L^{(k)}_{n}\}_{n=0}^\infty$ spans a dense
subspace of  the Hilbert space
\begin{equation}
  \label{eq:Wkdef}
  \cH_k := \rL^2([0,\infty),W_k dz),\quad 
  W_k(z) := z^k e^{-z}  ,
\end{equation}
and forms an orthogonal polynomial family satisfying
\begin{equation}
  \label{eq:Lnortho}
  \int_0^\infty L^{(k)}_{n} L^{(k)}_j W_k(z) dz =
  \begin{cases}
    \frac{\Gamma(k+n+1)}{n!} & n=j\\
    0 & n\neq j\,.
  \end{cases}
\end{equation}

The {\it two-step exceptional Laguerre polynomials} are now defined in
the following way.  Let $k$ be real and $0\leq m_1 < m_2$ integers.
Set
\begin{equation}
  \label{eq:elldef}
  \ell=m_1+m_2-1
\end{equation}
and define
\begin{align}
  \label{eq:etamdef}
  \eta_{12} &= \cW[\eta_1,\eta_2], \text{ where } \eta_a =
  L^{(-k-2)}_{m_a},\quad a=1,2,
\end{align} and where $\cW$ denotes the Wronskian operator.  Consider now the second-order linear differential operator $\hcL$ defined by 
\begin{align}
  \label{eq:hLdef}
  \hcL_k[y] &= zy''+(k+1-z)y'+\frac{2z}{\eta_{12}}\big( \eta_{12}'(y-y') +
    \cW[\eta_1',\eta_2']y\big).
\end{align}
Let $ n\geq \ell$ be an integer. Set
\begin{align}
  C&:= \left((m_1-m_2)(k+2-m_1+n)(k+2-m_2+n)\right)^{-1}
  \intertext{and let $ \hL_{n}$ be the polynomial defined by}
  \label{eq:hLndef}
  \hL_{n} :=\hL^{(k,m_1,m_2)}_{n} &= C z^{-k}\,
  \cW[\eta_1,\eta_2,z^{k+2} L^{(k+2)}_{n-\ell}],\qquad n\geq \ell\\
  \label{eq:hLnorm}
  &= \frac{(-z)^{n} }{(n-\ell)!\, m_1! \,m_2!} + \text{lower
    degree terms}.
\end{align}
For a lighter notation, we shall use $\hL_n$ instead of
$\hL^{(k,m_1,m_2)}_{n}$ throughout the paper. Whenever the symbol $\hL_n$ is
used on its own, the rest of the parameters $k\in\mathbb R$, $m_1,m_2\in\mathbb
N$ are understood.

The main results of our paper are the following.
\begin{theorem}
  \label{prop:hLeqn}
  The above-defined $\hL_n$ is the unique polynomial solution of the
  differential equation 
 \begin{equation}\label{eigen}
  \hcL_k[\hL_n] = (2\ell-n) \hL_n,\quad n\geq \ell
  \end{equation}subject
  to the normalization \eqref{eq:hLnorm}.
\end{theorem}
\begin{theorem} 
  \label{prop:2steportho}
  If $k> m_2-2$, then $\eta_{12}(z)>0$ for $z\geq 0$. Furthermore, the
  above defined polynomials $\{\hL_n\}_{n=\ell}^\infty$ span a dense
  subspace of the Hilbert space
  \begin{equation}
    \label{eq:hWdef}
    \cH_{k,m_1,m_2} := \rL^2([0,\infty),
    \hW_{k,m_1,m_2} dz),\quad  \hW_{k,m_1,m_2}(z) := z^k \eta_{12}^{-2} e^{-z}
  \end{equation}
  and satisfy the following orthogonality relations:
  \begin{equation}
    \label{eq:hL2ortho}
    \int_0^\infty \hL_n \hL_j \hW_{k,m_1,m_2}\, dz =
    \begin{cases}
      \displaystyle
      \frac{C\,\Gamma(k+n+3-\ell)}{(m_1-m_2)\,(n-\ell)!}&
      n=j\\  
      0 & n\neq j\,.
    \end{cases}
  \end{equation}
\end{theorem}

If $m_1=0, m_2=1$, then by well-known identities for the Laguerre
polynomials
\[ \hL^{(k,0,1)}_n(z) = \frac{z^2 L^{(k+2)}_n{}''(z) + (k+2)(2z
  L^{(k+2)}_n{}'(z) + (k+1) L^{(k+2)}_n(z))}{(k+2+n)(k+1+n)} =
L^{(k)}_n(z).\] Therefore, the above-defined polynomials generalize
the classical ones.  If $m_1=0, m_2=m+1$, where $m\geq 1$, then
\[\frac{C}{k+2+n-m}\,\hL^{(k,0,m+1)}_{n} = z^{-k} \cW[L^{(-k-1)}_m,z^{k+1}
L^{(k+1)}_n] = L^{\rm{II}}_{k,m,n},\quad n\geq m,\] where the latter
are the recently introduced exceptional Laguerre polynomials of type
II \cite{Sasaki-Odake1,GKM10}.  Therefore, the newly defined
polynomials $\hL_n$ also generalize and extend the previously
described class of exceptional Laguerre polynomials. This is natural
since for $m_1=0$ one of the factorizing functions becomes a constant
and therefore we are dealing with a 1-step Darboux transformation. The
two families of $X_\ell$-Laguerre exceptional polynomials of arbitrary
codimension were shown to be obtainable by means of a 1-step Darboux
transformation from the classical ones \cite{GKM10}.

\section{Rational factorizations} \label{RatFact}
This Section is devoted to the proof of Theorem \ref{prop:hLeqn}, characterizing
the two-step exceptional Laguerre polynomials $\hL_n $ as the unique normalized
polynomial solutions of the differential equation (\ref{eigen}).

Let \[T[y] = p(z) y''+q(z) y' + r(z) y\] be a second-order differential
operator with rational coefficients $p(z), q(z)$ and $r(z)$.
\begin{defi}
A \emph{rational factorization} of $T[y]$ is an operator identity of the form 
\[T= BA + \tlambda\] where $A[y], B[y]$ are first-order differential operators
with rational coefficients.
\end{defi}
We call \[\hT = AB + \tlambda\] the \textit{partner
operator} corresponding to the above rational factorization. Note that the
following formal intertwining relations therefore hold by construction:
\begin{equation}
  \label{eq:intertwiners}
  AT = \hT A,\qquad TB = B \hT.
\end{equation}
Given a rational factorization, let
$\phi(z)$ be the unique (up to constant multiple) solution of $A[\phi]
= 0$.  Observe that $w(z):= \phi'(z)/\phi(z)$ is rational and that
$T[\phi] = \tlambda\phi$.  For this reason we  will call $\phi(z)$
a {\it quasi-rational} factorization function.  Also note that the following
identities hold for $A$ and $B$:
\begin{align}
\label{eq:Aydef}
&A[y] = \frac{b(z)}{\phi}\, \cW[\phi,y] = b(z)\big(y'-w(z) y\big)\\
  \label{eq:Bydef}
  &B[y] = \frac{\hat{b}(z)}{\hphi}\, \cW[\hphi,y] = \hb(z)\big(y'-\hw(z) y\big),
\end{align}
where $b(z)$ is a rational function called the {\it factorization
  gauge} \cite{GKM10}, and
  \[\hat{b}  = \frac{p}{b},\quad \hat{w} = -w-\frac{q}{p}+\frac{b'}{b},\quad
\hat{\phi} =
  \exp\left(\int \hat{w}\,dz\right).
\]
The partner operator has the form
\begin{equation}
  \label{eq:hTdef}
  \hT[y] = p(z) y'' + \hq(z) y' +\hr(z) y
\end{equation}
where
\begin{equation}
  \label{eq:hqdef}
   \hq = q+p'-\frac{2pb'}{b},\quad \hr = -p(\hw'+\hw^2)-\hq \hw + \tlambda.
\end{equation}
Let us also define measures $W dz$ and $\hW dz$ by setting
\[ W = p^{-1 } \exp \left(\int \frac{q}{p}\, dz\right),\quad \hW =
\frac{{\hb}}{b}\,W\,.\]
The above defined operators $A$ and $-B$ are formally adjoint of each other
with respect to these measures, in the sense that:
\begin{equation}
  \label{eq:ABadjoint}  \int A[f] g\, \hW dz + \int B[g] f \, W dz = \hb W f g.
\end{equation}

It is important to note that a rational factorization is just special case of the well-known
Darboux transformation for Schr\"odinger operators, albeit expressed relative to a general
coordinate and gauge.  Indeed, the operator $T[\phi]$ can be related to a
Schr\"odinger operator
\begin{equation}
  \label{eq:Hdef}
  H[\psi] = -\psi'' + U(x) \psi
\end{equation}
by means of a
change of variable $z=\zeta(x)$ satisfying
\begin{equation}
  \label{eq:pzetarel}
  p(\zeta(x)) = \zeta'(x)^2  
\end{equation}
and a gauge transformation  
\begin{gather}
  \label{eq:gaugexform}
  -(\mu\, T[\phi])\circ \zeta = H[\psi],\\
  \label{eq:psidef}
  \psi= (\mu  \phi)\circ \zeta,   \\
  \label{eq:mudef}
  \mu = \exp\int \!\frac{2q-p'}{4p}dz,\\ 
  \label{eq:Urrel}
  U = -(r\circ \zeta)+ \frac{(\mu\circ\zeta)''}{(\mu\circ \zeta)}
\end{gather}
In the physical variable and gauge, the above factorization
corresponds to 
\[ H = (\partial_x + u)(-\partial_x + u)-\tlambda,\quad
u(x)=\frac{\psi'(x)}{\psi(x)},\] and is related to the factorization function
$\phi(z)$ by \eqref{eq:gaugexform}.  The partner operator in the
physical gauge is given by
\begin{equation}
  \label{eq:hHdef}
  \hH = (-\partial_x+u)(\partial_x+u)-\tlambda = -\partial_{xx} + \hU
\end{equation}
where the partner
potential is given by the well-known Crum potential formula
\begin{equation}\label{Crum}
  \hU = U - 2 \partial_{xx} \log \psi.
\end{equation}


Just as the Darboux transformation can be iterated to obtain the more
general Darboux-Crum transformation, rational factorizations can be
iterated and give rise to higher order intertwining relations.
Indeed, we shall see that the operator $\hL[y]$ defined in \eqref{eq:hLdef} is obtained
from the classical Laguerre operator $L[y]$ by means of a 2-step
factorization.  We describe this construction in more detail, and
thereby furnish a proof of Proposition \ref{prop:hLeqn}.

Fix $k, m_1, m_2$ as above and let
\begin{equation}
  \label{eq:T0def}
  T_0 = \cL_{k+2}
\end{equation}
be the ordinary Laguerre operator. The quasi-rational functions
\[\phi_i := z^{-k-2}\eta_i,\quad i=1,2\]  are
factorization functions of $T_0$ with eigenvalues $\tlambda_i = k+2-m_i$
\cite[Section 6.1]{erdelyi}.  Consider the rational factorization of $T_0$
corresponding to the factorizing function $\phi_1$ and factorizing gauge
\begin{equation}
  \label{eq:b1def}
  b_1 = z \eta_1
\end{equation}
which gives
\begin{equation}
  \label{eq:T0B1A1}
  T_0 = B_1 A_1+\tlambda_1, 
\end{equation}
where by equations  \eqref{eq:Aydef} and \eqref{eq:Bydef} it is clear that
\begin{equation}
  \label{eq:A1B1def}
  A_1[y] = \frac{b_1}{\phi_1}\, \cW[\phi_1,y],\qquad B_1[y] =
\frac{y'-y}{\eta_1}.
\end{equation}
Let 
\begin{equation}
  \label{eq:T1A1B1}
  T_1 := \hat T_0= A_1B_1+\tlambda_1
\end{equation}
be the corresponding partner operator.  Note now that
\[ \phi_{12} := A_1[\phi_2] = z^{k+3} \cW[\phi_1,\phi_2] = z^{-k-1} \eta_{12}
\] is a quasi-rational factorization function of $T_1$ with eigenvalue
$\tlambda_2 = k+2-m_2$.  This means that $\phi_{12}$ can be used as
factorization
function for a rational factorization of $T_1$, together with the factorization
gauge
\begin{equation}
  \label{eq:b12def}
  b_{12} = \frac{z \eta_{12}}{\eta_1}.
\end{equation}
We thus obtain the factorization
\begin{equation}
  \label{eq:T1B2A2}
  T_1 = B_2  A_2 + \tlambda_2
\end{equation}
where
\[ A_2[y] = \frac{b_{12}}{\phi_{12}}\, \cW[ \phi_{12},y] .\] 
We call $T_2$ the partner operator of this last factorization
\begin{equation}
  \label{eq:T2A2B2}
  T_2 := \hat T_1= A_2B_2+\tlambda_2
\end{equation}
and we observe that
\begin{equation}\label{A1A2}
 (A_2 A_1)[y] = b_{12} b_1
\frac{\cW[\phi_1,\phi_2,y]}{\cW[\phi_1,\phi_2]} = z^{-k}
\cW[\eta_1,\eta_2, z^{k+2} y].
\end{equation}
By construction, the following higher-order intertwining relation  holds 
\begin{equation}\label{higherint}
T_2 A_2 A_1 = A_2 A_1 T_0.\end{equation} 
From the definition of the $\hL_n$ polynomials in \eqref{eq:hLndef} and the
above expression \eqref{A1A2} we see that $\hL_n$ are eigenpolynomials of the
differential operator $T_2$:
\begin{equation}
  \label{eq:T21hLnrel}
  T_2[\hL_n]= (\ell-n) \hL_n.
\end{equation}
To conclude the proof of Theorem 1, we must establish that $T_2$ and $\hcL_k$
are related by
\begin{equation}
  \label{eq:hcLT2rel} 
  \hcL_k = T_2 +\ell.
\end{equation}
Let $q_i$ and $r_i$, $i=0,1,2$ be the coefficients of operators $T_i$ above:
\begin{equation}
  \label{eq:Tipqr}
  T_i[y] = z y'' + q_i y' + r_i \,y,\qquad i=0,1,2.
\end{equation}
From \eqref{Lagk} and \eqref{eq:T0def}, we see for instance that
\begin{align}
  \label{eq:q0def}
  q_0 &= k+3-z,\\
  \label{eq:r0def}
  r_0 &= 0\intertext{and hence by
 \eqref{eq:hqdef} \eqref{eq:b1def} \eqref{eq:b12def},}
  \label{eq:q2def}
  q_2 &= q_1+p'-2\,p\, \frac{b_{12}'}{b_{12}}\\ \nonumber
  &= q_0 + 2p'-2p\log(b_1 b_{12})'\\ \nonumber
  &=  1+k-z-2\,z\, \frac{\eta_{12}'}{\eta_{12}} .
\end{align}
which matches the first order part of \eqref{eq:hLdef}.
It remains to demonstrate that
\begin{equation}
  \label{eq:r2form}
  r_2 = \frac{2z}{\eta_{12}}(\eta_{12}'+\cW[\eta_1',\eta_2'])-\ell,
\end{equation}
and for this purpose it will be useful to compare these expressions in the
Schr\"odinger gauge by making use of the Crum potential formula.

Let $U_i(x), \mu_i(z),\; i=0,1,2$ be the potentials and gauge factors
corresponding to the operators $T_i$, as given by \eqref{eq:mudef}
-\eqref{eq:Urrel}.  In particular, by \eqref{eq:q2def},
\begin{align}
  \label{eq:mu0def}
  \mu_0 &= e^{-\frac{z}{2}} z^{\frac{5}4+\frac{k}2},\\
  \label{eq:mu2def}
  \mu_2 &= \frac{\mu_0}{z\eta_{12}} =  \frac{e^{-\frac{z}2}
z^{\frac{k}2+\frac{1}4}}{\eta_{12} }
\end{align}
where the latter follows  by \eqref{eq:q2def}.
Since $p(z)=z$, the required change of variables is
\begin{equation}
  \label{eq:zetaxdef}
  \zeta(x)=\frac{x^2}{4},
\end{equation}
and by the Crum potential formula \eqref{Crum}, we have
\begin{equation}
  \label{eq:U2def}
  U_2 = U_0 - 2 \partial_{xx} \log \cW[\psi_1,\psi_2],
\end{equation}
where $\psi_i  = (\mu_0\phi_i)\circ \zeta$.
It follows that
\begin{align}
  \label{eq:logcWrel}
  \log\cW[\psi_1,\psi_2] &= \big(\log\cW[\phi_1,\phi_2]+\frac{1}{2}\log p+
    2\log\mu_0\big)\circ \zeta\\ \nonumber
  &= \log (\eta_{12}\circ \zeta) - (1+k) \log \zeta - \zeta
\end{align}
By \eqref{eq:Urrel}, \eqref{eq:U2def} and \eqref{eq:r0def} we have
\[
r_2\circ\zeta 
  =  2 (\log \cW[\psi_1,\psi_2 ])''+
\frac{(\mu_2\circ\zeta)''}{(\mu_2\circ\zeta)}(x)-
\frac{(\mu_0\circ\zeta)''}{(\mu_0\circ\zeta)}(x).
\]
Applying the identity 
\[ f''/f - g''/g = (\log(f/g))'' + (\log(f/g))' (\log(fg))' \] with
$f= \mu_2\circ\zeta$ and $g=\mu_0\circ\zeta$ and relation
 \eqref{eq:mu2def} gives
\begin{align*}
  r_2\circ\zeta &= (\log (\eta_{12} \circ \zeta) -2\zeta-(2k+3)\log
  \zeta)'' +\\
  &\qquad +(\log\zeta+\log(\eta_{12}\circ\zeta))' (\log \zeta +
  \log(\eta_{12}\circ\zeta) -2\log(\mu_0\circ\zeta))'\\
  &= \frac{(\eta_{12}\circ\zeta)''}{(\eta_{12}\circ\zeta)}
  -(2\zeta+(2k+3)\log\zeta)''  +(\log\zeta)'(\log\zeta-
  2\log(\mu_0\circ\zeta))'\\
  &\qquad + 2(\log (\eta_{12}\circ\zeta))'
  (\log\zeta-(\mu_0\circ\zeta))'\\
  &=\frac{(\eta_{12}\circ\zeta)''}{(\eta_{12}\circ\zeta)} + 2(\log
  (\eta_{12}\circ\zeta))'
  (\log\zeta-(\mu_0\circ\zeta))'
\end{align*}
Now switching back to the algebraic variable gives
\begin{align}
  \nonumber
  r_2 &= \frac{(z \eta_{12}''+\frac{1}2\eta_{12}')} {\eta_{12}} +
  2z\left(\frac{1}z-\frac{\mu_0'}{\mu_0}\right)\frac{\eta_{12}'}{\eta_{12}} \\
  \label{eq:r2etarel}
  &= \frac{z\eta_{12}''}{\eta_{12}}+(z-k) \frac{\eta_{12}'}{\eta_{12}}.
\end{align}
Since $y=\eta_i,\; i=1,2$ are solutions of the Laguerre differential
equation, 
\begin{equation}
  \label{eq:etailaguerre}
  zy''-(1+k+z)y'+m_i y=0,
\end{equation}
we have
\[ z\eta_{12}'' -(k+z) \eta_{12}' -2z
\cW[\eta_1',\eta_2']+\ell \eta_{12} = 0,\]
which together with \eqref{eq:r2etarel} implies
\eqref{eq:r2form}, the required conclusion. We have thus established that the
polynomials $\hL_n$ defined by \eqref{eq:hLndef} are obtained from the
associated Laguerre polynomials $L^{(k)}_{n}$  by use of a two-step rational
factorization.
The proof of Theorem 1 is achieved by showing that the differential equation
$\hcL_k$ in \eqref{eq:hLdef} satisfied by the polynomials $\hL_n$ is essentially
the partner of Laguerre's operator, corresponding to the iterated Darboux
transformation.
It is important to stress that the two factorizing functions of
the iterated rational factorization:
\[ \phi_1=z^{-k-2}\,L_{m_1}^{(-k-2)}(z),\qquad
\phi_2=z^{-k-2}\,L_{m_2}^{(-k-2)}(z)\]
are indexed by two integers $m_1$ and $m_2$ that can be freely chosen. This
fact should be put in comparison with the restricted choice of factorizing
functions in the original Crum method \cite{crum} or its modification by Adler \cite{adler}. The main difference lies also in the fact that in the former constructions the Darboux transformations are state-deleting, while in this paper the factorizing functions correspond to isospectral transformations.

\section{Proofs of completeness and orthogonality
properties}\label{completeness}
\begin{lemma}
  \label{lemma1}
  Let $m_2>m_1\geq0$ be integers and $k>m_2-2$ a real number. Then,
  $\eta_{12}(z)>0$ for $z\geq 0$.
\end{lemma}
\noindent\textit{Proof.} Since $y=\eta_i,\; i=1,2$ satisfies \eqref{eq:etailaguerre},  $\eta_{12}$ satisfies the inhomogeneous
first-order differential equation
\[
z\eta_{12}'-(k+1+z)\eta_{12}+(m_{2}-m_{1})\eta_1 \eta_2=0.
\]
This gives
\[
\eta_{12}(z)=(m_2-m_1) e^{z}z^{1+k}\int_z^{\infty}e^{-t} t^{-2-k}\eta_1(t) \eta_2(t)
dt.
\] If $k>m-2$, then by Theorem 6.73 of \cite{Sz},
$L^{(-k-2)}_{m}(z)>0$ for $z\geq 0$. The desired conclusion follows
immediately.

Let $\cP$ denote the vector space of univariate polynomials, and
$\cP_n$ the subspace of polynomials of degree $\leq n$.  Let $\ell =
m_1+m_2-1$ and let $\cE_n\subset\cP_{n+\ell}$ denote the span of
$\hL_{\ell},\ldots, \hL_{n+\ell}$.  Let $z_i,\; i=1,\ldots,\ell$
denote the zeroes of $\eta_{12}(z)$.
\begin{lemma}\label{Lemma2}
A polynomial $y\in
  \cP_{n+\ell}$ belongs to $\cE_n$ if and only if
\begin{equation}\label{eq:constr}
  -2z_i y'(z_i) +\left(z_i-k+\frac{z_i\eta_{12}''(z_i)}{\eta_{12}'(z_i)}\right)
\,y(z_i) =0,\qquad 
  i=1,\ldots,\ell
\end{equation}
\end{lemma}
\noindent
\textit{Proof.} Suppose that $y\in \cE_n$. By equations
\eqref{eq:q2def} \eqref{eq:r2etarel} of the previous section, 
\[ \Res\Big\{(-2z y' \frac{\eta_{12}'}{\eta_{12}} +y\,\frac{(z-k)\eta_{12}'+
  z\eta_{12}''}{\eta_{12}}\,,\,z =z_i\Big\} = 0,\qquad i=1,\ldots,
  \ell. \] 
These are precisely the relations \eqref{eq:constr}.  Since the
codimension of $\cE_n$ in $\cP_{n+\ell}$ is exactly $\ell$, the
constraints \eqref{eq:constr} define the subspace $\cE_n$, as claimed.

\begin{lemma}\label{Lemma3}
  Let $\eta(z)$ be a polynomial such that $\eta(z)\neq 0$ for $z\geq
  0$. If $k>-1$, then the subspace $\eta \cP := \{ \eta(z) p(z) \mid
  p\in \cP \}$ is dense in $\cH_k$.
\end{lemma}
\noindent \textit{Proof.} Since $\cP$ is dense in $\cH_k$  \cite{Sz},
it suffices to prove that $\cP$ is contained in the $\rL^2$ closure of
$\eta\cP$. Let $p\in\cP$ and $\epsilon>0$ be given.  We seek a
polynomial $q\in \cP$ such that $\|p-q\eta\|_k \leq \epsilon$.  To
this end let $z_0 = 1+\max \{ |z| : \eta(z) = 0 \}$ and set
\[ r(z)=\begin{cases}
  \frac{p(z-z_0)}{\eta(z-z_0)}\quad & z\geq z_0,\\
  0 \quad &\text{otherwise}
\end{cases}
\]
Set $l = k+2\deg \eta$ and observe that 
\[ z^k|\eta(z)|^2< (z+z_0)^l,\quad z\geq 0.\] Since $1/|\eta(z)|$ is
bounded for $z\geq 0$, we have
\[ \| r\|_l=\int_{z_0}^\infty
\left|\frac{p(z-z_0)}{\eta(z-z_0)}\right|^2 z^l e^{-z} dz =
\E^{-z_0}\int_0^\infty
\left|\frac{p(z)}{\eta(z)}\right|^2 (z+z_0)^l \E^{-z}dz <\infty\] Since
$\cP$ is dense in
$\cH_\ell$, there exists a $\tq\in \cP$ such that $\| r-\tq\|_l \leq
\epsilon$.  Hence, $q(z) = \tq(z+z_0)$ is the desired polynomial
because
\begin{align*}
  \| p-q\eta\|_k &= \int_0^\infty |r(z+z_0)-\tq(z+z_0)|^2 \,|\eta(z)|^2
  z^k \E^{-z}
  dz\\
  &< \int_0^\infty |r(z+z_0)-\tq(z+z_0)|^2 \, |z+z_0|^l\, \E^{-z} dz\\
  &< \E^{z_0}\int_0^\infty |r(z)-\tq(z)|^2\, z^l \E^{-z} dz\\
  &< \epsilon
\end{align*}
This concludes the proof of Lemma~\ref{Lemma3}. 

We now use Lemma \ref{Lemma2} and Lemma \ref{Lemma3} to prove Theorem \ref{prop:2steportho}.  The positivity of
$\eta_{12}(z)$ was established in Lemma \ref{lemma1}. It therefore follows that the measure $\hW_{k,m_1,m_2}(z) dz$ has finite moments. Let us now prove that the set of all 2-step Darboux exceptional Laguerre polynomials
\[\cE=\bigcup_{n} \cE_{n} = \text{span}\{ \hat L_n\}_{n=\ell}^\infty \]
 is dense
in $\cH_{k,m_1,m_2}$.  Let $f\in \cH_{k,m_1,m_2}$ and $\epsilon>0$ be given.
Set $\tf = f/\eta_{12}$ and note that $\tf\in \cH_k$ because $\| f
\|_{k,m_1,m_2} = \| \tf \|_k$.  Hence, by Lemma \ref{Lemma3}, there
exists a polynomial $p\in \cP$ such that
\[ \| f-p \eta_{12}^2\|_{k,m_1,m_2} = \int^\infty_0 |
\tf(z)-\eta_{12}(z) p(z)|^2 z^k e^{-z} dz \leq \epsilon.\] By Lemma
\ref{Lemma2}, $p\eta^2_{12}\in \cE$, which establishes the claim.

To conclude the proof of Theorem \ref{prop:2steportho} let us prove the orthogonality \eqref{eq:hL2ortho} of the exceptional polynomials.  Let $A_i[y], B_i[y],\;
i=1,2$ be the first-order operators defined in the preceding section.
For $f,g\in \cP$, by \eqref{eq:intertwiners}\eqref{eq:ABadjoint}
\eqref{eq:T0B1A1} \eqref{eq:T1B2A2}
\begin{align*}
  \left< A_2A_1f,A_2A_1g\right>_{k,m_1,m_2} &= \left< B_1 B_2 A_2
    A_1 f, g\right>_{k+2} \\
  &= \left< B_1 (T_1+\tlambda_2) A_1 f , g\right>_{k+2}\\
  &= \left< (T_0+ \tlambda_2)(T_0 + \tlambda_1) f, g\right>_{k+2}
\end{align*}
The desired conclusions follow by \eqref{eq:Lneigenrelation} and
\eqref{eq:Lnortho}.

\section{Summary and conclusions}

We have shown how the iterated Darboux or Darboux-Crum transformation can be used to construct new families of exceptional orthogonal polynomials, thus enlarging the class previously described in the literature.
Due to the many properties in common with classical orthogonal polynomials, a systematic classification of all the exceptional families is an important goal, and the Darboux-Crum construction described in this paper seems  an appropriate tool to achieve it.

In the Schr\"odinger gauge and physical variable, this construction gives rise to new exactly solvable potentials in quantum mechanics.

\vskip.4cm
\noindent\textbf{Acknowledgments}

\smallskip
The research of DGU was supported in part by MICINN-FEDER grant MTM2009-06973
and CUR-DIUE grant 2009SGR859. The research of NK was supported in part by NSERC
grant RGPIN 105490-2004. The research of RM was supported in part by NSERC grant
RGPIN-228057-2004.

\end{document}